\documentclass{aa}
\usepackage{graphics,epsf,epsfig,graphicx}

\begin{document}

\title{Very Luminous Carbon Stars in the Outer Disk of the Triangulum
  Spiral Galaxy}

\author{David L. Block \inst{1}, Kenneth C. Freeman \inst{2},
        Thomas H. Jarrett \inst{3}, Iv\^anio Puerari \inst{4},
        Guy Worthey \inst{5}, Fran\c coise Combes \inst{6},
        Robert Groess \inst{1}}

\offprints{D.L. Block \email{block@cam.wits.ac.za}}
\institute{
School of Computational and Applied Mathematics,
University of Witwatersrand, Private Bag 3, WITS 2050, South Africa
\and
Mount Stromlo and Siding Spring
Observatories, Research School of Astronomy and Astrophysics, Australian
National University, Australia
\and
Infrared Processing and Analysis Centre,
100-22, CALTECH, 770 South Wilson Ave, Pasadena, CA 91125, USA
\and
Instituto Nacional de Astrof\'\i sica,
Optica y Electr\'onica,
Calle Luis Enrique Erro 1, 72840 Tonantzintla, Puebla, M\'exico
\and
Washington State University, 1245 Webster Hall,
Pullman, WA 99163-2814, USA
\and
Observatoire de Paris, LERMA, 61 Av. de
l'Observatoire, F-75014, Paris, France }
\date{Received XX XX, 2004; accepted XX XX, 2004}
\authorrunning{Block et al.}
\titlerunning{Very Luminous Carbon Stars in the Outer Disk of the
  Triangulum Galaxy}

\abstract{Stars with masses in the range from about $1.3$ to $3.5~
M_\odot$ pass through an evolutionary stage where they become carbon
stars. In this stage, which lasts a few Myr, these stars are extremely
luminous pulsating giants. They are so luminous in the near-infrared
that just a few of them can double the integrated luminosity of
intermediate-age (0.6 to 2 Gyr) Magellanic Cloud clusters at 2.2
microns. Astronomers routinely use such near-infrared observations to
minimize the effects of dust extinction, but it is precisely in this
band that carbon stars can contribute hugely.  The actual contribution
of carbon stars to the outer disk light of evolving spiral galaxies has not
previously been morphologically investigated.  Here we report new and
very deep near-IR images of the Triangulum spiral galaxy M33 $= $ NGC 598, 
delineating
spectacular arcs of carbon stars in its outer regions. It is these arcs
which dominate the near-infrared $m=2$ Fourier spectra of M33. We
present near-infrared photometry with the Hale 5--m
reflector, and propose that the arcs are the signature of accretion of
low metallicity
gas in the outer disk of M33.  
%which
%demonstrate that carbon stars are indeed abundant.  Exploiting this
%result, we show that these highly evolved stars can skew the luminosity
%function of intermediate-brightness galaxies by up to $80$\%. The
%physical significance is that the luminosity function measures the
%fraction of baryons that have been converted into stars since the big
%bang.  The uncorrected effect of carbon stars on the luminosity
%function will cause us to overestimate both the number of stars and the
%average star formation rate throughout the history of the universe.
%Furthermore, most galaxies will pass through a first carbon star phase,
%when the universe was only about 10\% of its present age and carbon
%stars were rampant. At this phase, the galaxies have a redshift $\sim
%4$, so the dominant output from the carbon stars will move into the
%mid-infrared where these stars could double the observed galactic
%flux.  Instruments like the Mid-Infrared Instrument on board the JWST
%will be ideal for imaging these galaxies in their (rest-frame) 2.2
%micron band.
}

\maketitle
\section{Introduction}
Carbon stars are thermally pulsing asymptotic giant branch (TP--AGB)
stars with ages between about $0.6$ to $2$ Gyr. They are observed in
the intermediate-age globular clusters of the Large and Small
Magellanic Clouds.  Although the average number of carbon stars per
intermediate age cluster in the Magellanic Clouds is only about 2.5
(Persson et al. \cite{persson83}), they radiate such a copious amount of near-IR light
that they contribute about 50\% of the bolometric luminosity of the
cluster (Marigo et al. \cite{marigo03}, Maraston \cite{maraston98}).  From stellar population
synthesis studies, the near-IR $K$-band ($2.2\mu$) luminosity is enhanced
in galaxies containing a significant intermediate-age stellar
population (Mouhcine and Lan\c con \cite{mouhcine03}) by up to a
factor $2$.  
In this Letter,
we present new deep near-IR images of the nearby spiral galaxy M33
which illustrate the dramatic effect that an intermediate-age
population can make to the near-IR light distribution in spiral
systems. 
%We also discuss some of the observed and potential
%consequences of this near-IR enhancement from TP-AGB stars.

\begin{figure*}
\vspace{15.0cm}
%\special{psfile=block_fig1.eps
%hscale=80 vscale=80 angle=0
%hoffset=0 voffset=0}
\caption{{\bf (a).} Reaching six times deeper than 2MASS is this
$JHK_s$ image of M33, with a simple ellipsoid model subtracted.
A gargantuan plume-like ring of red stars
stretches in a swath (up to 5$'$ in width)
for over 120 degrees, commencing at $\sim$ 14$'$  north
of the galaxy centre. A fainter counterpart is found to the south. The
surface brightness of the northern arc is only $\sim$ 20-21 mag
arcsec$^{-2}$ at 2.2 $\mu m$, which explains why it has hitherto not
been discussed in previous near-infrared studies.
An ellipsoidal swath is selected
to pass through the plumes; that sector passing through
the prominent northern plume is color-coded red.
{\bf (b).} As in (a), but  deprojected, beautifully 
reveals the plumes as
well as the inner arms. The northern plume is labelled A to B.
{\bf (c).} A deep 2MASS $H-$band image of M33 (non-deprojected),
reveals the northern arc (arrowed) and other red arcs in the outer
disk of M33, but their presence is best
shown by subtracting away an axisymmetric component. 
The arcs appear to form a ring. It is tempting to
liken this ring to the one recently reported in the outer
disk of our Milky Way (Ibata et al. \cite{ibata}).
{\bf (d).} Fourier spectra generated from the deprojected near-infrared $K_s$
mosaic of M33. The dominant $m=2$ mode in M33 does not arise from the inner
pair of spiral arms seen in Figure 1a, but from the giant outer ring or arcs
of stars identified in this study: a prominent plume in the north, and a
fainter counterpart to the south. While M33 in the optical shows ten spiral 
arms (Sandage and Humpheys \cite{sandage}), the galaxy in the near-infrared has only
two low-order modes: $m=1$ and $m=2$.} 
\end{figure*}

Astronomers routinely use near-IR observations of spiral galaxies
to minimize the effects of dust extinction, but the actual contribution
of carbon stars to the near-IR light of evolving galaxies remains
poorly quantified.  Little has changed observationally since Aaronson
noted the importance of the asymptotic giant branch for understanding
the stellar content of nearer galaxies (Aaronson \cite{aaronson}).  Disks of
galaxies appear to form from the inside out (Block et al. \cite{blocketal02}).
Relative to the inner regions of spiral galaxies, the mean ages of
the outer regions are known to be somewhat younger and more
metal-poor (Bell and de Jong \cite{belldejong00}). We can therefore expect the contribution
from the intermediate-age stars to be stronger in these outer
regions, and the near-IR surface brightness of the outer disk will
be preferentially enhanced by the presence of TP--AGB stars.  This
brightening of the outer disk in the near-IR may well contribute
to the apparent sharp radial truncation observed in the disks of
many spiral systems (Kregel et al. \cite{kregeletal02}).

In this context, we present new near-infrared images of the Local Group
spiral M33 from a special set of 2MASS observations. The integration
time for these images was increased by a factor of six, extending
approximately 1 mag deeper than the standard survey.  
Three mosaics are constructed for M33, corresponding to the $J$
      (1.2 $\mu m$), 
      $H$ (1.6$\mu m$), and `$K$-short', $K_s$ (2.2$\mu m$) bands. 
At $K_s$, the image resolves larger-area features 
as faint as  $\sim$22.5 mag arcsec$^{-2}$.

Foreground Milky
Way stars were removed statistically, using the $J-K_s$ histogram of two
control star-fields to the East and West of M33 (but at the same
galactic latitude) as a template.  Of the $\sim$7000 sources in
the original image,  $\sim$2300 were removed as foreground stars.

\section{Analysis}

M33 is famous for an optically bright pair of spiral
arms, but surprisingly, it is not
these two arms which dominate the near-infrared $m=2$ Fourier spectra.  
The near-infrared images reveal remarkable arcs of red stars in the
outer disk of M33, spanning $120^\circ$ in azimuth angle.  The northern
arc is dominant although a very faint southern counterpart arc, forming
a partial ring, can also be seen (Figure 1).  A hint of the
northern swath can be seen in the study by Regan and Vogel (1994). 
Fourier analysis of the
light distribution (Figure 1d) shows that the dominant $m=2$ peak
corresponds to the giant arcs.  The pitch angle of the dominant $m=2$
structure is $\sim$  58 degrees.  

\begin{figure}
\vspace{7.0cm}
%\special{psfile=block_fig2.eps
%hscale=25 vscale=25 angle=0
%hoffset=35 voffset=0}
\caption{An overlay of the integrated cluster
light in the Magellanic Clouds (from
Figure 4 of Persson et al. \cite{persson83}),
with the integrated colors for plume stars 
in the NE sector (green dots) and SW sector (red dots) of M33.
The open symbols principally show the Searle, Wilkinson and
Bagnuolo (\cite{searle80}) (SWB) Magellanic
cluster types I, II and III; these
do not contain carbon stars.
The filled symbols are SWB types V, VI and VII. The
red Magellanic carbon-star-bearing
clusters  lie above and to the right of the
black dashed lines, in the Magellanic ``IR-enhanced'' cluster
regime identified by Persson et al. (\cite{persson83});
these are predominantly SWB types V and VI. Red bars show our
1-sigma uncertainty in the color
for each point in this figure; the black error bars are
from the Persson et al. (\cite{persson83}) analysis. 
The thin blue dotted and solid lines show the Bessell \&
Brett (\cite{bessel}) tracks for giants and dwarfs respectively.
The overlay provides strong evidence for a carbon star population
in the M33 plumes.}
\end{figure}

\begin{figure}
\vspace{7cm}
%\special{psfile=block_fig3.eps
%hscale=30 vscale=30 angle=0
%hoffset=35 voffset=25}
\caption{A $J-H$, $H-Ks$ color-color diagram contrasting the nuclear
  colors with those of  northern inner
  spiral arm and the NE and SW plume populations.The 
 evolved giant and main-sequence dwarf tracks are shown 
      with green dashed and solid lines, following Bessell and
  Brett (\cite{bessel}).  }
\end{figure}

\begin{figure}
\vspace{7cm}
%\special{psfile=block_fig4.eps
%hscale=25 vscale=25 angle=0
%hoffset=35 voffset=25}
\caption{Observations with the Hale 5m reflector, through a
   sector of the northern red arc. The sample has been cleaned of 
      all low detection sources, with a $S/N < 10$. 
      The extinction reddening vector is indicated with the red 
      arrow.  The evolved giant and main-sequence dwarf tracks are
      from Bessell and Brett (1998).  
      In this color-color plot, no foreground MW stars 
      were statistically removed, as in the other color-color plot. 
      Here foreground stars appear with blue colors 
      ($H-K_{s} < 0.2$  and $J-H < 0.8$; $J-K_{s} < 1.0$), 
 while most of the M33 
      sources have colors that are redder than 
      $H-K_{s}$ = 0.4 and $J-H$ = 0.9 ($J-K_{s}$ = 1.3) mag. 
The color uncertainties 
      for the red sources are less than 10\%.} 
\end{figure}

The very red color of the arcs is not due to dust.  Wilson (\cite{wilson91})
derived E($B-V$) = 0.3 $\pm$ 0.1 mag, which includes both foreground
(Milky Way) and internal M33 extinction.  The $K-$band extinction
A($K$) is estimated $\sim$ 0.09 mag (Wilson, 1991).

The color of the northern arc extends to $J-K_s > 1.1$.  Very old M
giants of solar abundance can reach $J-K_s \sim 1$ (see e.g.  Figure 2
in Bessell and Brett \cite{bessel}), and even redder if they are
super-metal-rich (Frogel and Whitford \cite{frogel}). However, as reviewed by Pagel and
Edmunds (\cite{pagel}), there is a strong radial abundance
gradient (Searle \cite{searle}) in M33 ($-0.09 \pm 0.02$ dex/kpc in O/H); the
outer regions are relatively metal-poor, and solar abundance is reached
only within the inner 1$''$ of M33.  In regions of lower abundance, the
giant branch is bluer.  If stars with $J-K_s > 1$ are found in the outer
low-metallicity regions, they cannot be M-giants.  Figure 2 presents
the integrated colors of the plumes in the NE and SW sector of M33, overlayed
with the integrated colors of the clusters in the Magellanic
Clouds (Persson et al. \cite{persson83}).  Figures 2 and 3 
provide strong evidence for a
carbon star population in the extended M33 plumes.  We conclude that
the near-IR arcs are probably dominated by a population of very red
carbon stars.

 Only integrated fluxes 
      with 
      a S/N greater than 10 are shown in Figure 3; hence, the accuracy of the 
      photometry is comparable (to within a factor of two) to 
    that of Persson et 
      al (1983). 
      Note that individual M33 stars are not 
    resolved by 2MASS (unlike the 
      Palomar 
      \hbox{observations} as discussed below), 
 and so the colors best represent the 
 `local'  area (300 sq. 
      arcsec) of 
      the targeted regions.  Individual AGB and carbon stars will `redden' the 
      $J-H$ and $H-Ks$ colors, but the overall effect is 
    mitigated by the stellar 
      population 
      in the local region.  Consequently, the colors represent a 
    lower limit to 
      the color of the \hbox{evolved} population.

Why are the Fourier spectra not dominated by the inner, very robust
spiral arms? In a spiral galaxy with a sustained star formation
history, in the first few Myr red supergiants will dominate the
infrared
light. 
In the 
     0.5-2 Gyr 
      range carbon stars dominate, especially in a somewhat metal-poor 
      regime. 
      After 
      this time, the carbon stars die and ordinary RGB/AGB stars dominate. In 
      the 
      first 2 Gyr, 
      the C-stars clearly dominate, as accreted gas in the outer disk 
 is fresh and not 
      yet 
      well mixed 
      by M33's near-infrared oval/bar. In the inner disk, 
   there presumably are older populations present that 
      dominate 
      in mass, and therefore in light, and the $Ks$-band enhancement
      of the TP-AGB carbon stars is weakened.

One of us (THJ) recently imaged a section of the northern 
      plume of 
      M33 
      with the 5m-Hale reflector, using the 2048$\times$2048 
array near-infrared 
      camera 
      WIRC (see Figure 4). The field of view is $8.5'\times8.5'$ with 
      0.25 arcsec pixels,  
%The seeing FWHM is 0.8$''$ in J, and 0.7 
%      $''$ 
%      in 
%      the $K_{s}$ band.  The telescope was centered at 
%      01h34m28.1s,  +30d54m00s (J2000). 
%      The total $JHK_{s}$ integration time was $\sim$ 9 minutes, 
      reaching a limiting surface brightness at $K_{s}$ of 23.7 
      mag per arcsec$^{-2}$.
%  (at 1$\sigma$, the RMS is 22.21 mag
%      arcsec$^{-2}$ per pixel; 4$\times$4 pixels gives 1 square
%      arsec),
 %     The point source photometry 
%      $S/N=10$ limits are 19.0, 18.0, 16.9 mag in $JHK_s$, respectively. 
%      Figure 4 shows the photometric 
%      $J-H$ 
%      and $H-K_{s}$ colors for sources 
%      detected 
%      in the 76 arcmin$^{2}$ field.  
      A plethora of stars is seen toward the C-Star regime in the 
      upper right corner of the figure, with colors exceeding 
      $J-K_{s} \sim  1.8$  mag.

 In their modelling of the HI envelope, 
  Corbelli and Schneider (1997) find that the phase changes
    at a radius of $\sim$ 20 arcmin, which  is precisely the
    outer domain of the arcs reported in this study. 
  It seems highly plausible therefore, that fresh, low-metallicity gas is
  being fed to the host
 galaxy M33, via external accretion. The outer disk from which mass is
  accreted
 is inclined to the inner disk of M33, and has a different angular
     momentum.

The accretion of gas 
      from gaseous  filaments is expected to be asymmetric, since 
      it 
      comes from one side 
      only -  gas is being accreted and only later becomes bound and 
      finds its way 
      to 
      circular 
      orbits. It takes a dynamical time to relax. In the present epoch, we 
      expect the 
      signature of accreted gas (the external arcs) to be asymmetric. 
  We believe that it is the signature of gas flowing
  inward and accreting at $\sim$ 20 arcmin, from which
  the very red,
 and relatively metal-poor stars have been formed. This
is also fully consistent with the recent observations elucidated
by Tiede et al. (2004).

\section{Conclusion}

It is tempting to liken the ring in M33 to the one recently reported
in the outer disk of our Milky Way
(e.g. Ibata et al. \cite{ibata}).
For the Milky Way ring,  
the rotation period is $\sim$ 600 Myr. For M33, the rotation
time-scale at the radius of the ring would be $\sim$ 240 Myr.
Any carbon-bearing 
clusters of age 0.6 Gyr would have only undergone $\sim$ 2
orbits.

%Carbon stars may make an important contribution to the IR luminosity
%of high-redshift galaxies.  Using the Large Magellanic Cloud as a
%guide, carbon stars are produced in large numbers between ages of
%about 0.6 and 2 Gyr.  Therefore galaxies that undergo a burst of
%star formation will have their IR light boosted 0.6 Gyr later, and
%this extra light will die away after about 2 Gyr. We know that star
%formation in the early universe 
%(Bouwens et al. \cite{bouwens03}) was already proceeding
%at redshifts $z \sim 6$.  There may be an epoch starting about 0.6
%Gyr after the onset of star formation in the universe (i.e. at
%redshifts $z \sim 4$) characterized by large numbers of carbon
%stars. The dominant output from the carbon stars will be redshifted
%into the mid-infrared where these stars could double the observed
%galactic flux. Instruments like the Mid-Infrared Instrument on
%board the JWST are needed to image these galaxies in their (rest-frame)
%2.2 micron band.

Our discovery of a dominant
carbon-star-bearing stellar population in the outer regions of a
spiral galaxy (M33)
shows how imperative it is to use stellar population synthesis 
models  which include carbon
stars (e.g. Mouhcine and Lan\c con \cite{mouhcine2002}), 
when analysing the intregated light of spiral galaxies. 
The outer arcs in M33 are not revealed on standard near-infrared surveys,
such as 2MASS, because their integrated surface brightness is fainter
than the 2MASS threshold.  
It is a
sobering thought that it takes a groundbased 
class 4-5m telescope to resolve  
individual carbon stars in the outer disk of M33, our second closest
spiral.

\acknowledgements{We thank the anonymous referee for his/her
  insightful comments. We are very grateful to Mike Bessell, Matthew
Colless, Simon Driver, Ariane Lan\c con and Roberto de Propris for
advice.  DLB thanks the Research School of Astronomy \& Astrophysics for
visitor status and support.  KCF and IVP thank 
the Anglo American Chairman's Fund 
and the University of the Witwatersrand for hospitality.}

\end{document}